\pdfoutput=1
\documentclass[prl,twocolumn,preprintnumbers,nofootinbib,superscriptaddress,10pt]{revtex4-1}
\usepackage[dvips]{graphicx}

\usepackage{amsmath,amsthm,amssymb,multirow,psfrag}
\usepackage{epsfig}
\usepackage{color}
\usepackage{hyperref}

\graphicspath{{./Figures/}}

\begin{document}

%%%%%%%%%%%new definitions: 
\def\lsim{\mathrel{\rlap{\lower4pt\hbox{\hskip1pt$\sim$}}
  \raise1pt\hbox{$<$}}}
\def\gsim{\mathrel{\rlap{\lower4pt\hbox{\hskip1pt$\sim$}}
  \raise1pt\hbox{$>$}}}
\newcommand{\vev}[1]{ \left\langle {#1} \right\rangle }
\newcommand{\bra}[1]{ \langle {#1} | }
\newcommand{\ket}[1]{ | {#1} \rangle }
\newcommand{\ev}{ {\rm eV} }
\newcommand{\kev}{{\rm keV}}
\newcommand{\Mev}{{\rm MeV}}
\newcommand{\Gev}{{\rm GeV}}
\newcommand{\Tev}{{\rm TeV}}
\newcommand{\mw}{$M_{W}$}
\newcommand{\Ft}{F_{T}}
\newcommand{\Zparity}{\mathbb{Z}_2}
\newcommand{\BLambda}{\boldsymbol{\lambda}}
\newcommand{\met}{\;\not\!\!\!{E}_T}
\newcommand{\beq}{\begin{equation}}
\newcommand{\eeq}{\end{equation}}
\newcommand{\bea}{\begin{eqnarray}}
\newcommand{\eea}{\end{eqnarray}}
\newcommand{\nn}{\nonumber}
\newcommand{\gev}{{\mathrm GeV}}
\newcommand{\hc}{\mathrm{h.c.}}
\newcommand{\eps}{\epsilon}
\newcommand{\bwt}{\begin{widetext}}
\newcommand{\ewt}{\end{widetext}}
\newcommand{\draftnote}[1]{{\bf\color{blue} #1}}
\newcommand{\vecsigma}{ \mbox{\boldmath $\sigma$}}

\newcommand{\Mpl}{M_{\rm Pl}}
\newcommand{\TRH}{T_{\rm RH}}
\newcommand{\TCMB}{T_{\rm CMB}}
\newcommand{\xiend}{\xi_{\rm end}}
\newcommand{\Hend}{H_{\rm end}}
\newcommand{\aend}{a_{\rm end}}

\newcommand{\cO}{{\cal O}}
\newcommand{\cL}{{\cal L}}
\newcommand{\cM}{{\cal M}}

%References  
\newcommand{\fref}[1]{Fig.~\ref{fig:#1}} 
\newcommand{\eref}[1]{Eq.~\eqref{eq:#1}} 
\newcommand{\aref}[1]{Appendix~\ref{app:#1}}
\newcommand{\sref}[1]{Section~\ref{sec:#1}}
\newcommand{\tref}[1]{Table~\ref{tab:#1}}

%%%%%%%%%%%%%%%%%%%%%%%%
\newcommand{\LU}[1]{\textcolor{blue}{[LU: #1]}}

%%%%%%%%%%%%%%%%%%%%%%%%%%

%%%%%%%%%%%%%%%%%%%%%%%%%%%%%%%%%%
\title{\Large{{\bf Vector dark matter production at the end of inflation}}}

%%%%%%%%%%%%%%%%%%%%%%%%%%%%%%%%%%%%%

\author{\bf {Mar Bastero-Gil}}
\email{mbg@ugr.es}
\affiliation{\normalsize\it
Departamento de F\'{i}sica Te\'{o}rica y del Cosmos and CAFPE,~Universidad de Granada, Campus de Fuentenueva, E-18071 Granada, Spain}

\author{\bf{Jose Santiago}}
\email{jsantiago@ugr.es}
\affiliation{\normalsize\it
Departamento de F\'{i}sica Te\'{o}rica y del Cosmos and CAFPE,~Universidad de Granada, Campus de Fuentenueva, E-18071 Granada, Spain}

\author{\bf{Lorenzo Ubaldi}}
\email{ubaldi.physics@gmail.com}
\affiliation{\normalsize\it
SISSA International School for Advanced Studies,Via Bonomea 265, 34136 Trieste, Italy}
\affiliation{\normalsize\it
INFN - Sezione di Trieste, Via Bonomea 265, 34136, Trieste, Italy}

\author{\bf{Roberto Vega-Morales}}
\email{rvegamorales@ugr.es\\}
\affiliation{\normalsize\it
Departssamento de F\'{i}sica Te\'{o}rica y del Cosmos and CAFPE,~Universidad de Granada, Campus de Fuentenueva, E-18071 Granada, Spain}

\preprint{UG-FT 328-18,~CAFPE 198-18,~SISSA 44/2018/FISI}
\begin{abstract}
It has been shown that the longitudinal mode of a massive vector boson can be produced by inflationary fluctuations and account for the dark matter content of the Universe.~In this work we examine the possibility of instead producing the transverse mode via the coupling $\phi F \tilde F$ between the inflaton and the vector field strength.~Such a coupling leads to a tachyonic instability and exponential production of one transverse polarization of the vector field, reaching its maximum near the end of inflation.~At production the mass is negligible and the vectors add up coherently to form a dark electromagnetic field.~As the Universe expands, the energy density of the dark electromagnetic field then redshifts like radiation until its wavelength stretches to beyond its Compton wavelength.~After this point the vectors become non-relativistic and their energy density redshifts like matter.~We show that these polarized transverse vectors can account for the observed dark matter relic density in the mass range $\mu$eV to hundreds of GeV.~We also find that the tachyonic production mechanism of the transverse mode can accommodate larger vector masses and lower Hubble scales of inflation compared to the production mechanism for the longitudinal mode via inflationary fluctuations.

\end{abstract}
\maketitle

%%%%%%%%%%%%%%%%%%%%%%%%%%%%%%%%%%
\section{Introduction} \label{sec:intro} 

The nature of dark matter still remains a mystery.~While most of the experimental effort in the past was aimed at detecting weakly interacting massive particles (WIMPs), the lack of observation has necessitated new theoretical ideas and proposed search strategies to cover as many alternatives as possible~\cite{Bertone:2018xtm}.~In this context, it is important to explore new production mechanisms for dark matter and assess their impact on search strategies. 

Dark photons are among the alternative dark matter candidates which have received increasing attention in recent years.~They can be produced via the misalignment mechanism like axions~\cite{Nelson:2011sf}, though only if they couple non-minimally to gravity~\cite{Arias:2012az}.~They can also be produced via inflationary fluctuations~\cite{Graham:2015rva} which is particularly interesting since it only relies on the single condition that the Stueckelberg mass of the vector\,\footnote{We use `dark photon' and `vector' interchangeably throughout.}~is a good effective description at least up to the Hubble scale of inflation.~If this is the case, the longitudinal mode of the vector is inevitably produced by inflation and constitutes a viable dark matter candidate for a large range of values of its mass and Hubble scale of inflation.

In this work we examine the possibility of producing the \emph{transverse} component of the vector during inflation.~We exploit the coupling $\phi F\tilde F$ between the inflaton and the dark photon which leads to the exponential production of \emph{one} transverse mode of the dark photon via tachyonic instability.~This generates a dark electromagnetic (DEM) field at the end of inflation with a large energy density that redshifts like radiation.~As the Universe expands and the typical wavelength of the DEM stretches to the point where it reaches the Compton wavelength, the dark photon then becomes non relativistic and redshifts like matter, as necessary for a viable dark matter candidate.~We show that there is a large region of parameter space in which this production mechanism dominates over that of the longitudinal mode~\cite{Graham:2015rva}, thus allowing for the transverse mode of the vector to account for the dark matter relic abundance we observe today.

We do not specify the form of the inflaton potential in this study as the only properties relevant to our analysis are the Hubble scale of inflation and the presence of the $\phi F\tilde F$ coupling.~The latter is generically present in models of axion inflation~\cite{Pajer:2013fsa} for which exponential production of gauge bosons has been explored in many studies.~These studies have examined various possibilities including; the production of primordial magnetic fields~\cite{Field:1998hi,Anber:2006xt,Adshead:2016iae} as a dissipation mechanism which allows for inflation on a steep potential~\cite{Anber:2009ua}, the generation of primordial curvature perturbations and non-gaussianities~\cite{Barnaby:2011vw, Linde:2012bt}, as well as the production of primordial black holes~\cite{Linde:2012bt, Garcia-Bellido:2016dkw} and gravitational waves~\cite{Anber:2012du, Garcia-Bellido:2016dkw,Adshead:2018doq}.~All of these possibilities utilize effects of the axion-photon coupling in the initial or middle e-folds of inflation.~In contrast, the mechanism proposed here focuses on the very last few e-folds of inflation where the production of the gauge quanta is largest.

We also perform an initial exploration of the dark photon power spectrum at the end of inflation which is needed as input for the cosmological evolution of the energy density and a more precise quantification of the final dark matter relic abundance.~A mored detailed investigation of the cosmic evolution of the power spectrum and energy density will be presented in followup work~\cite{followup}.

Below we first review the mechanism for tachyonic production of the gauge field (dark photon in our case) during inflation and examine the final energy density.~We then estimate the final relic abundance of dark photons showing how it depends on the model parameters and detailing the assumptions and constraints of our analysis.~We also briefly discuss our mechanism in the context of quantum gravity and consider some conjectured constraints on the dark photon mass while emphasizing that our mechanism is insensitive to whether the vector mass is of the Stueckelberg or Higgs-ed type.~We then conclude and discuss possibilities for followup work.

%%%%%%%%%%%%%%%%%%%%%%%%%%%%%%%%%%
\section{Dark photon production} \label{sec:prod}

In this section we outline the mechanism for generating dark photon dark matter via tachyonic instability during inflation.~Much of the discussion presented here mirrors discussions found in studies of exponential production of massless gauge bosons via the $\phi F \tilde F$ coupling~\cite{Anber:2009ua,Barnaby:2010vf,Barnaby:2011vw,Barnaby:2011qe,Anber:2012du,Meerburg:2012id,Pajer:2013fsa} in the context of models of natural inflation~\cite{Freese:1990rb}.

%%%%%%%%%%%%%%%%%%%%%%%%%%%%%%%%%%
\subsection{Tachyonic production during slow-roll inflation} \label{sec:tachy} 

We start with the action for an inflaton with potential $V(\phi)$ coupled to a massive neutral vector boson,
\bea
S  &=& - \int d^4x \, 
\sqrt{-g} \Big[\frac{1}{2} \partial_\mu \phi \partial^\mu \phi + V(\phi) \nonumber \\ 
&+&~\frac{1}{4} F_{\mu\nu} F^{\mu\nu} + \frac{1}{2} m^2 A_\mu A^\mu + \frac{\alpha}{4 f} \phi F_{\mu\nu} \tilde F^{\mu\nu} \Big] \, ,
%, \label{eq:action} 
\label{eq:Lag}
\eea
where $\phi$ is the inflaton field that drives inflation, $A_\mu$ is the dark photon field, $F_{\mu\nu}= \partial_\mu A_\nu - \partial_\nu A_\mu$ is the field strength, and $\tilde F^{\mu\nu} = \epsilon^{\mu\nu\alpha\beta} F_{\alpha\beta}/2$ with $\epsilon^{\mu\nu\alpha\beta}$ the completely antisymmetric tensor.~We use the Friedmann-Robertson-Walker (FRW) metric with $ds^2 = - dt^2 + a^2(t) d\vec{x}^2$ and the convention $\epsilon^{0123} = 1/\sqrt{-g}$.~The vector boson mass $m$ can be zero or non-zero during inflation and can be of a Stueckelberg or `Higgsed' type (we comment on this point later).~As long as it is smaller than the Hubble scale at the end of inflation the vector mass has negligible effects on the tachyonic production mechanism.~We also assume negligible kinetic mixing between the visible and dark photons which is also not necessary for the production mechanism\,\footnote{Allowing for small kinetic mixing does not destroy the production mechanism and can lead to interesting phenomenology~\cite{Graham:2015rva,Alexander:2016aln}, but for simplicity we take it to be small enough to be neglected.}.~We do not specify the form of the inflaton potential as the only relevant properties needed for our discussion are:~$(i)$ the scale of $V(\phi)$ which sets the Hubble scale during inflation,
\bea
H = \frac{\sqrt{V(\phi)}}{\sqrt{3} \Mpl} \, ,
\eea
and $(ii)$ the coupling of the inflaton to $F\tilde F$ responsible for exponential production of only one polarization of the transverse modes.~Such a coupling is generically present in models of natural inflation, where $\phi$ is a pseudoscalar (odd under parity) axion-like field subject to a shift symmetry.~For this reason, this class of models provides a well motivated theoretical framework for the mechanism presented here.~However, since the dynamics of the mechanism do not depend on whether $\phi$ is an axion or not, $\phi$ can be a generic scalar (or function of $\phi$~\cite{Barnaby:2011qe}), as long as it drives inflation and couples to $F \tilde F$.~Note that we do not need to impose that the Lagrangian respects parity so $\phi$ could also be parity even.~As a numerical example to illustrate the mechanism, we will consider below a potential of the form $V(\phi) = \lambda^4 \phi^4 / 4$, while axion like periodic potentials and others are examined in~\cite{followup}. 

We quantize the vector field by expanding in terms of creation and annihilation and their mode functions,
\bea
\hat{\vec A}(\vec x,t) &=& 
\sum_{\lambda = \pm,L} \int \frac{d^3k}{(2\pi)^3} 
e^{i \vec k \cdot \vec x} \ \vec \epsilon_\lambda(\vec k) \\
&\times&[A_\lambda(k,t) a_\lambda(\vec k) + A_\lambda(k,t)^\ast a_\lambda^\dagger(-\vec k)] , \nonumber
\eea
where we include transverse and longitudinal polarizations in the sum and the creation and annihilation operators satisfy the commutation relation,
\bea
\Big[a_\lambda(\vec k),\,a_\lambda^\dagger(\vec k^\prime) \Big] = (2\pi)^3 
\delta_{\lambda\lambda^\prime}\,\delta^3(\vec k - \vec k^\prime) .
\eea
The mode functions obey the equations of motion derived from~\eref{Lag} which in Fourier space read~\cite{Anber:2009ua,Barnaby:2010vf},
\bea
\ddot \phi &+& 3 H \dot\phi + V'  = 
\frac{\alpha}{f} \langle F \tilde F \rangle  , \label{eq:EOMphi} \\
\ddot A_\pm &+& H \dot A_\pm + \left( \frac{k^2}{a^2} \pm \frac{k}{a} \frac{\alpha\dot\phi}{f} + m^2  \right) A_\pm  = 0 \, , \label{eq:AEOMT} \\
\ddot A_L &+& \frac{3 k^2 + a^2 m^2}{k^2 + a^2 m^2} H \dot A_L + \left( \frac{k^2}{a^2} + m^2 \right) A_L = 0 \, ,
\label{eq:AEOML}
\eea
where we have also included the inflaton equation of motion.~The overdots denote derivatives with respect to physical time $t$ and $k \equiv |\vec k|$ is the magnitude of the comoving momentum.~We consider only the spatially homogeneous zero momentum mode ($k=0$) of the inflaton.~We have separated the three degrees of freedom of the dark photon into transverse and longitudinal components, $\vec A_T$ and $A_L$ respectively, where $\vec k \cdot \vec A = k A_L$ and $\vec k \cdot \vec A_T = 0$, and we have written the transverse component in terms of the two helicities, $\vec A_T = \vec\epsilon_+ A_+ + \vec\epsilon_- A_-$.~The equation of motion for $A_L$ corresponds to the one derived in~\cite{Graham:2015rva} which is not affected by the presence of the coupling $\phi F \tilde F$.~Hence the longitudinal mode in this model can be produced via inflationary fluctuations, as described in~\cite{Graham:2015rva}, and contribute to the final dark photon relic density. 

In what follows we concentrate on the equation of motion of the transverse modes in~\eref{AEOMT}.~It is customary to introduce the dimensionless parameter,
\bea \label{eq:xidef}
\xi \equiv \frac{\alpha \dot \phi}{2H f} = \sqrt{\frac{\epsilon}{2}} \frac{\alpha}{f} \Mpl  \, ,
\eea
where $\epsilon \equiv - \dot H/H^2$ and for single field inflation we have,
\bea\label{eq:slowroll}
|\dot\phi| \approx  V'/3H,~~ 
\epsilon =  \frac{\dot \phi^2}{2H^2 M_{Pl}^2} .
\eea
It is useful to rewrite the equation of motion in terms of conformal time $\tau$ defined as $a d\tau = dt$, 
\bea
\label{eq:AEOMtau}
\Big[ \frac{\partial^{2}}{\partial\tau^{2}} + k^{2} 
\pm 2\,k\,\frac{\xi}{\tau} + \frac{\bar{m}^2}{\tau^2}
\Big] A_\pm(k, \tau)
=0 \, ,
\eea
where we have defined $\bar m \equiv m/H$ and used the fact that during inflation ($\tau < 0$) we have $\tau \simeq -\frac{1}{a H}$.~Using the convention $\dot\phi > 0$ we have $\xi > 0$, implying that \emph{only} the mode $A_+$ experiences tachyonic enhancement when,
\bea \label{eq:Omegatach}
\Omega^2(k,\tau) 
= -\omega^2(k,\tau) = -k^2 - 2k \frac{\xi}{\tau} - \frac{\bar{m}^2}{\tau^2} > 0 \, . 
\eea
On the other hand, the opposite polarization $A_-$~does not have tachyonic modes and is therefore neglected.~Treating $\xi$ as a constant, a good approximation during slow-roll, one can solve~\eref{AEOMtau} analytically~\cite{Meerburg:2012id} in terms of Whittaker functions.~In the massless case, from \eref{Omegatach} we see that the tachyonic modes are those with $-k\tau < 2 \xi$, and that $\Omega^2$ is maximal at $(-k\tau)_{\rm tachyonic} = \xi$.~We are interested in the limit of small vector mass, 
\bea \label{mtach}
\frac{m}{H} \equiv \bar m \ll (-k\tau)_{\rm tachyonic} = \xi \, .
\eea
For $\bar m \gg \xi$ the mass term quenches the tachyonic production~\cite{Meerburg:2012id}.~Neglecting the dark photon mass, the full analytic solution to~\eref{AEOMtau} can be found in terms of Coulomb functions and is normalized by the requirement that the gauge field is initially in the Bunch-Davies vacuum~\cite{Meerburg:2012id,Adshead:2015pva,Domcke:2018eki} (in the sub-horizon limit $-k\tau \to \infty$),
\bea\label{eq:bdvac}
\lim_{-k\tau\to \infty} A_\pm (k,\tau) = \frac{e^{-ik\tau}}{\sqrt{2k}} \equiv A_{\rm{BD}} \, .
\eea
We are interested in the regime $-k\tau < 2\xi$ ($k < 2 \xi a H$) where we start to produce tachyonic modes and the solution is very well approximated by~\cite{Barnaby:2011vw},
\bea \label{eq:Atachsol}
A_+(k,\tau) &\simeq & \sqrt{\frac{-2 \tau}{\pi}} e^{\pi \xi} K_1 \left[ 2 \sqrt{-2\xi k \tau} \right] \, , \\
-k\tau & < & 2\xi \, ,  \nonumber
\eea
where $K_1$ is a modified Bessel function of the second kind and the exponential enhancement is explicit.~From this one finds in the super-horizon limit $-k\tau\to 0$,
\bea \label{eq:ATP}
\lim_{-k\tau\to 0} A_+ (k,\tau) = \frac{e^{\pi \xi}}{2 \sqrt{2 \pi k \xi}} \equiv A_{\rm{TP}} \, ,
\eea
where TP stands for tachyonically produced.~Note that the amplitude of the gauge field $A_+(k,\tau)$ is maximal and equal to $A_{\rm TP}$ at $-k\tau \to 0$ and decreases for larger values of $-k\tau$.~To get further intuition, it is often useful to work with the solution obtained via the WKB approximation~\cite{Anber:2009ua, Barnaby:2011vw, Tangarife:2017rgl} which we can write as,
\bea \label{eq:AWKB}
A_+(k,\tau)_{\rm WKB} & \simeq & \frac{1}{\sqrt{2k}} \left( \frac{-k\tau}{2\xi} \right)^{1/4} e^{\pi \xi - 2 \sqrt{-2\xi k\tau}} \, , \\
\frac{1}{8\xi} & < &  -k\tau  < 2\xi \, , \nonumber
\eea
where the regime of validity is dictated by the adiabatic condition $|\Omega' / \Omega^2| \ll1$ with $\Omega' \equiv d\Omega/d\tau$.~In the range $1/(8\xi) < -k\tau < 2\xi$, $A_+(k,\tau)_{\rm WKB}$ approximates very well the solution obtained in~\eref{Atachsol} giving us intuition into the behavior of the modes around horizon crossing as they become exponentially enhanced.

%%%%%%%%%%%%%%%%%%%%%%%%%%%%%%%%%%%%%%%%%%%%%%%%%%%%
\subsection{Energy density of dark photons\\ at the end of inflation}

The modes which grow exponentially are those with physical momenta which satisfy,
\bea \label{eq:qtach}
q \equiv \frac{k}{a} < 2 \xi H \, .
\eea
In most models, which have roughly 60 e-folds of inflation, $\xi$ during the first few e-folds is constrained by CMB measurements~\cite{Barnaby:2010vf,Barnaby:2011vw} to be less than $\xi \lesssim 2.5$.~Here we are interested in $\xi$ at the end of inflation, $\xi_{\rm end}$, which is allowed to be significantly larger.~However, in order to neglect back reaction effects, we limit ourselves to $\xi_{\rm end} < \mathcal{O}(10)$ in the following.~See the discussion before~\eref{backreaction} for more detail.

From~\eref{qtach} we see that the tachyonic modes have at most $q \sim \mathcal{O}(20) H$, implying that their wavelength is roughly comparable to the size of the Hubble horizon as the modes become tachyonic.~Hence, within the horizon these modes add up coherently.~Given the very large occupation number due to the exponential enhancement in the solutions~\eref{Atachsol} and~\eref{AWKB}, these dark photons are well described by a classical dark electromagnetic field whose `electric' and `magnetic' components we label with $E$ and $B$ respectively.~Furthermore, since one helicity contributes exponentially more than the other, we are left with a \emph{polarized} dark electromagnetic field. 

To eventually compute the relic abundance we are interested in the energy density of the dark electromagnetic field.~This is well approximated~\cite{Anber:2009ua} by,
\bea
\rho_D &= & \frac{1}{2} \langle \vec E^2 + \vec B^2 \rangle \\
&=& \frac{1}{4\pi^2 a^4} \int_0^\infty dk \, k^2
\Big(
\left\vert \partial_\tau A_+ \right\vert^2 
+ k^2 \vert  A_+ \vert^2 \Big) 
\label{eq:rho} \\
&=& \frac{1}{2a^4} \int d\ln k \ \left(\mathcal{P}_{\partial_\tau A_+}(k,\tau) + k^2 \mathcal{P}_{A_+}(k,\tau)\right) \,, ~~~
\label{rhoPower}
\eea
where we have introduced the power spectrum,
\begin{align} \label{Pdef}
& \langle X(k,\tau)^* X(k,\tau) \rangle  \equiv  \frac{2\pi^2}{k^3}\mathcal{P}_X (k,\tau) \, , \\
& X = A_+ \ {\rm or} \ \ \partial_\tau A_+ \, .
\end{align}
We can also examine the quantity in the integrand,
\bea \label{eq:drho}
\frac{d\rho_D}{d \ln k} = \frac{1}{2a^4} \left(\mathcal{P}_{\partial_\tau A_+}(k,\tau) + k^2 \mathcal{P}_{A_+}(k,\tau)\right) \, ,
\eea
which provides us with information on which $k$ modes contribute most significantly to the energy density.~We are interested in finding the peak of $\frac{d\rho_D}{d \ln k}$ as a function of $k$ at the end of inflation when $\xi$ is the largest.~Using the WKB solution in~\eref{AWKB}, we can compute analytically the dark electric field contribution given by $\mathcal{P}_{\partial_\tau A_+}(k,\tau)$ as well as the dark magnetic field contribution $k^2 \mathcal{P}_{A_+}(k,\tau)$, where the former is always dominant over the latter.~We can then estimate $\rho_D$ during inflation,
\bea\label{eq:rhoDinf}
\rho_D 
\approx 10^{-4} \frac{H^4\, e^{2\pi \xi}}{\xi^3} \, .
\eea

The WKB as well as the other solutions in the previous section were derived under the assumption of constant $\xi$.~However, at the end of inflation $\xi$ increases and its variation $\dot\xi$ can be significant, depending on the specific model of inflation.~Thus, in this regime, treating $\xi$ as constant in the derivation is not a good approximation.~As the equations of motion can no longer be solved analytically with a time varying $\xi$, we turn to a numerical study in order to understand the detailed behavior.

We consider, as a concrete example, an inflaton with potential $V(\phi) = \lambda \phi^4/4$ and $\lambda = 10^{-14}$ which is the order of magnitude needed to obtain the correct amplitude for the primordial spectrum at CMB scales.~Inflation ends when the slow-roll parameter $\epsilon$ is equal to 1.~We solve numerically the coupled equations of motion \eref{EOMphi} and \eref{AEOMT} neglecting the back-reaction term $\alpha / f \langle F \tilde F \rangle$ on the right hand side of \eref{EOMphi}.~We consider three different values of the coupling $\alpha \Mpl / f$, corresponding to $\xiend = 3, 6, 9$, with $\alpha\Mpl / f = \sqrt{2}\, \xiend$ (see~\eref{xidef}).~For $\xiend = 3$ and 6 the back-reaction term $\alpha / f \langle F \tilde F \rangle$ is always negligible while for $\xiend=9$ it becomes relevant only at the very end of inflation in this specific example.

Going on to compute $\frac{d\rho_D}{d \ln k}$ at the end of inflation and normalizing it to the energy density of the inflaton at that point, $V(\phi_{\rm end}) = 3 \Hend^2 \Mpl^2$ with $\phi_{\rm end} = 2 \sqrt{2} \Mpl$, we show the result in the lower panel of \fref{PSRH}.~The striking feature is that $\frac{d\rho_D}{d \ln k}$ is sharply peaked at $k \approx a_{\rm end}H_{\rm end}$ for the different values of $\xiend$.~This is not just a consequence of the specific model of inflation under consideration as we also find similar results for a $V(\phi) \propto \phi^2$ potential.~We can understand this generic feature by considering the conformal diagram in the upper panel of \fref{PSRH}.
%%%%%
\begin{figure}
~~~~~~~~
\includegraphics[scale=.45]{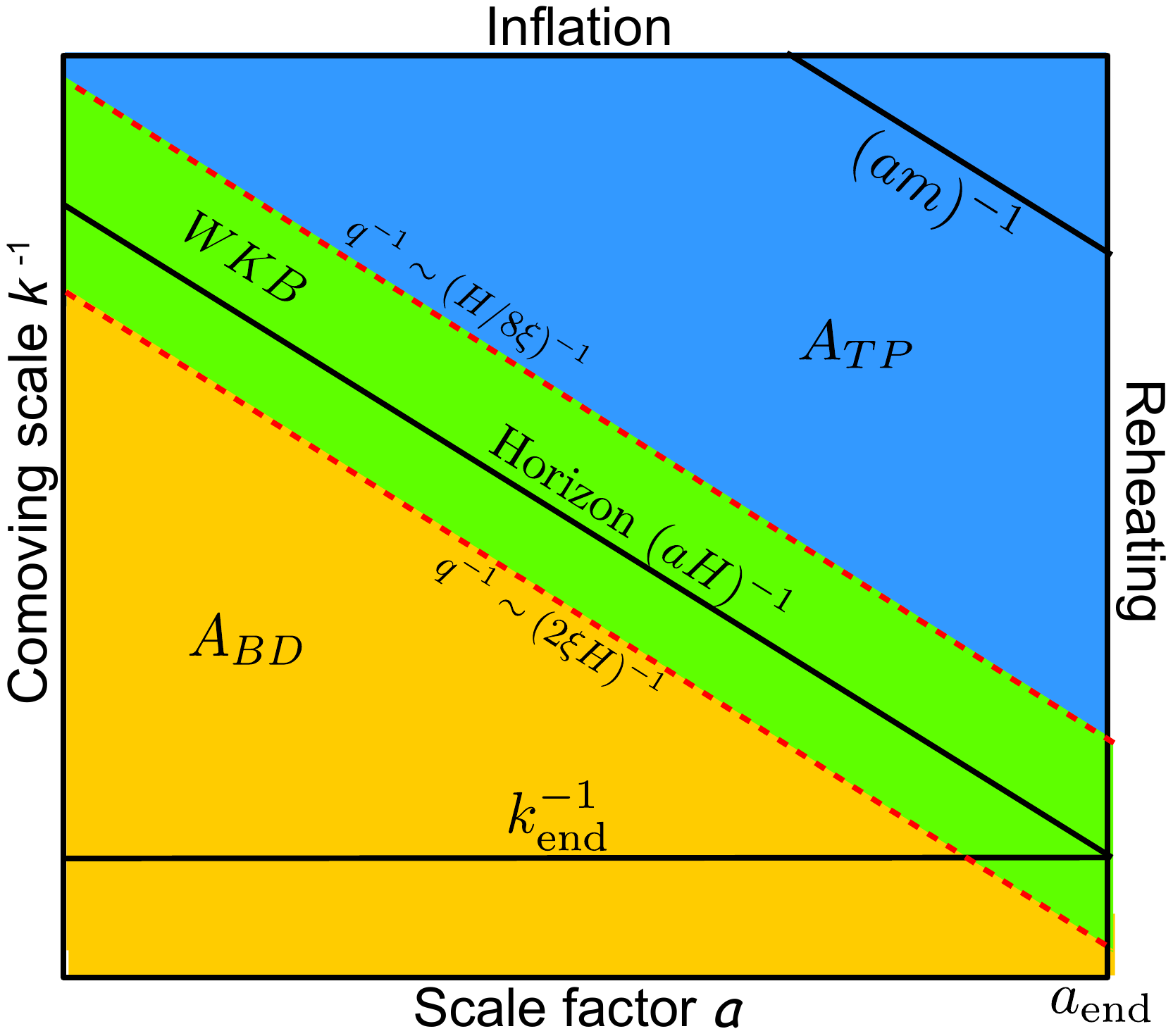}\\
\vspace{1cm}
\includegraphics[scale=.66]{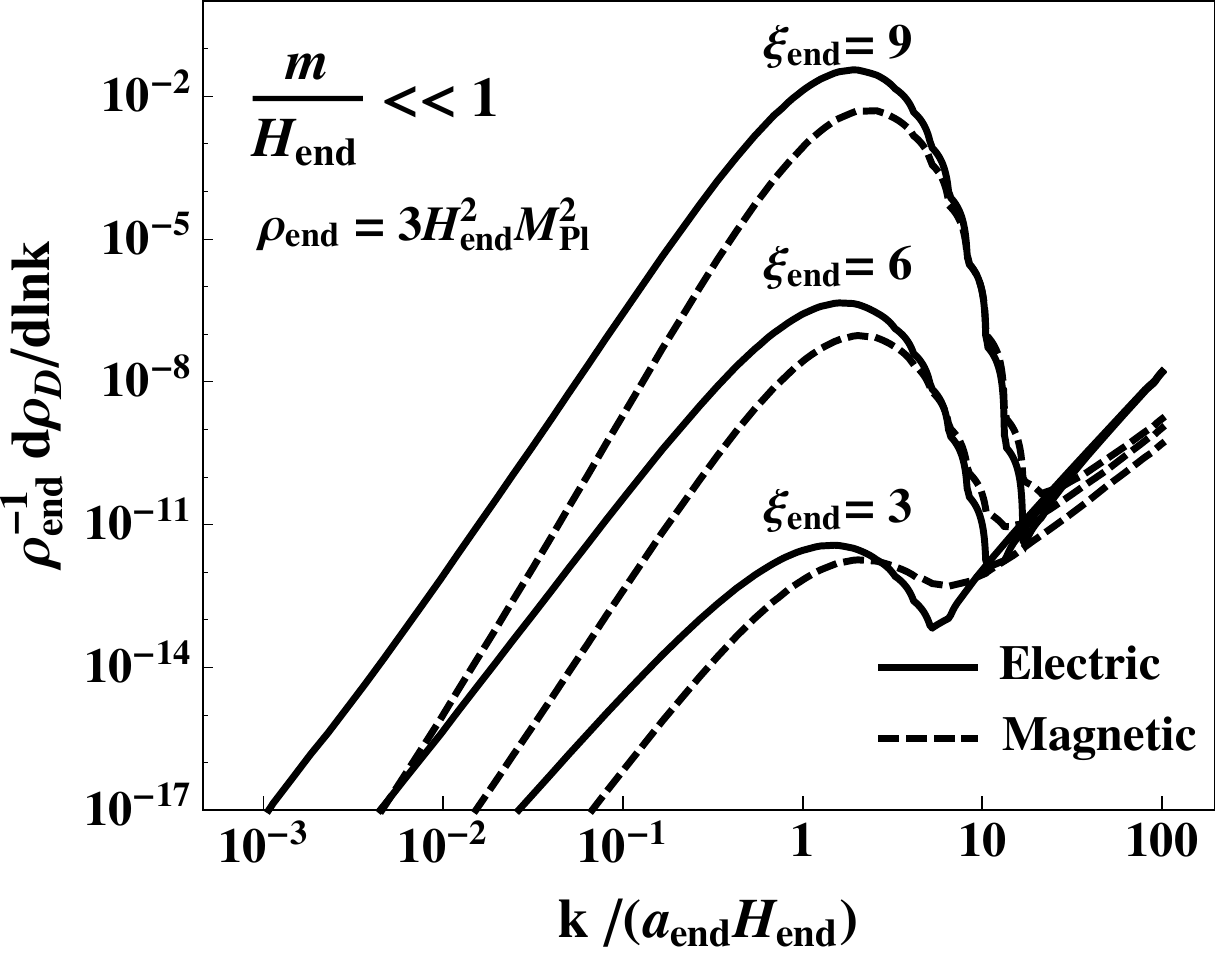}~~~
\caption{{\bf Top:~}Conformal diagram showing the cosmological evolution of comoving length scales $k^{-1}$ as a function of scale factor $a$ (or conformal time).~{\bf Bottom:}~The quantity $\frac{d\rho_D}{d \ln k}$ defined in~\eref{drho}, normalized to the energy density of the inflaton at the end of inflation, as a function of $k/(a_{\rm end}H_{\rm end})$ for an inflaton potential $V(\phi) = \lambda \phi^4/4$ with $\lambda = 10^{-14}$.
}
\label{fig:PSRH}
\end{figure}
%%%%%

In the conformal diagram, time (or scale factor) increases from left to right while the comoving momentum $k$ increases from top to bottom.~Modes with lower $k$ (larger comoving scale) enter the green tachyonic band early on during inflation when $\xi$ is still small.~Hence, these modes experience less exponential enhancement compared to modes at larger $k$ which enter the green band later during inflation when $\xi$ has grown larger.~This leads to a significant suppression at large comoving scale compared to $k^{-1} = (a_{\rm end}H_{\rm end})^{-1}$ and ensures that too much power is not generated.~Thus the usual scale invariant primordial spectrum is obtained at CMB scales as required by observation.~At comoving scales $k^{-1} < (a_{\rm end}H_{\rm end})^{-1}$ the comoving momentum quickly becomes larger than $k = -2\xiend / \tau_{\rm end}$, after which there is no tachyonic enhancement leading to a suppression and thus a peaked structure at $k \approx a_{\rm end}H_{\rm end}$.~We also see in~\fref{PSRH} that the dark electric field component dominates over the dark magnetic field one, as expected.

A more in depth study of the early and late time power spectrum is left to ongoing work~\cite{followup}.

%%%%%%%%%%%%%%%%%%%%%%%%%%%%%%%%%%
%%%%%%%%%%%%%%%%%%%%%%%%%%%%%%%%%%
%%%%%%%%%%%%%%%%%%%%%%%%%%%%%%%%%%
\section{Dark matter relic abundance} \label{sec:dmrelic}

In this section we go on to estimate the present day dark matter relic abundance.~During inflation when $\phi$ dominates the energy density we have for the inflaton,
\bea \label{eq:rhoI}
\rho_I = V(\phi) = 3 H^2 \Mpl^2 \, .
\eea
Once inflation ends, a fraction of $\rho_I$ is transferred to radiation to reheat the Universe while another fraction goes into the dark electromagnetic field.~We can write the radiation energy density in the visible sector as,
\bea \label{eq:rhoRH}
\rho_R (T_{\rm RH}) = \frac{\pi^2}{30}g_*(\TRH) \TRH^4 = \epsilon_R^4 3 H^2 \Mpl^2 \, ,
\eea
which defines the reheating temperature,
\bea \label{eq:TRH}
\TRH = \epsilon_R\, \left( \frac{90}{\pi^2 g_*(\TRH)} \right)^{1/4} \sqrt{H \Mpl} \, .
\eea
Here $g_*(\TRH)$ denotes the number of relativistic degrees of freedom.~We take $g_*(\TRH) \sim 100$ as we consider reheating temperatures above the weak scale.~The dimensionless parameter $\epsilon_R < 1$ parametrizes the fraction of the inflaton energy which goes into radiation and we work in the approximation of instantaneous reheating assuming it takes place as soon as the inflaton exits slow-roll.  

The Hubble parameter decreases from the beginning to the end of inflation.~We parametrize its final value as,
\bea \label{eq:Hend}
\Hend = \epsilon_H H \, ,
\eea
with $\epsilon_H$ a dimensionless parameter.~Typically the slow-roll parameter $\epsilon = - \frac{\dot H}{H^2}$ is $\mathcal{O}(10^{-2} - 10^{-3})$ during inflation and 1 at the end of inflation which translates into values of $\epsilon_H$ that are model dependent.~For most models of inflation, we expect $\epsilon_H$ to be in the range $10^{-3} < \epsilon_H < 10^{-1}$.~The parameter $\xi$, which controls the dark photon production and grows with $\dot\phi$, is largest at the end of inflation.~Thus, the largest contribution to the dark electromagnetic energy density comes from near the end of inflation, as confirmed in our numerical analysis shown in the power spectrum in~\fref{PSRH}.~Therefore, from \eref{rhoDinf} we can estimate the energy density of the dark photons produced at the end of inflation as,
\bea \label{eq:rhoD}
\rho_D(\TRH) = 10^{-4} \frac{\epsilon_H^4 H^4}{\xiend^3} e^{2\pi \xiend} \, .
\eea
Note that the parameter $\epsilon_H$ also effectively parametrizes what fraction of the available energy density goes into the dark electromagnetic field.~From~\fref{PSRH} we see that the tachyonic modes which give the largest contribution to $\rho_D(\TRH)$ have physical momentum,
\bea\label{eq:qRH}
q(\TRH) \equiv \frac{k}{\aend} \sim  \Hend \, .
\eea
At reheating we have $q(\TRH) \gg m$ and the physical momentum then redshifts as,
\bea
q(T) = q(\TRH) \frac{T}{\TRH} \, .
\eea
The dark photons become non relativistic at the temperature $\bar T$ defined by the condition,
\bea
q(\bar T) = m \, , 
\eea
which allows us to solve for $\bar{T}$ as,
\bea
\bar T = m  \left( \frac{90}{\pi^2 g_*(\TRH)} \right)^{1/4} \frac{\epsilon_R}{\epsilon_H}   \left( \frac{\Mpl}{H} \right)^{1/2} \, .
\label{eq:Tbar}
\eea
Above $\bar T$ the energy density $\rho_D$ redshifts like radiation, 
\bea \label{eq:rhoDrad}
\rho_D(\bar T) = \rho_D(\TRH) \left( \frac{\bar T}{\TRH} \right)^4 \, ,
\eea
while below $\bar T$ it redshifts like matter giving,
\bea \label{eq:rhoDmat}
\rho_D(\bar T) = \rho_D(T_0) \left( \frac{\bar T}{T_0}  \right)^3 \, ,
\eea
where $T_0 \approx 10^{-13}$ GeV is today's CMB temperature.~Combining~\eref{rhoD} and~\eref{Tbar}~-~\eref{rhoDmat}, we obtain the dark photon energy density today:
\bea
\rho_D(T_0) &=&  m \,T_0^3 \, \left( \frac{90}{\pi^2 g_*(\TRH)} \right)^{-3/4} 
\left( \frac{H}{\Mpl} \right)^{3/2}  \nonumber \\
 &\times& 10^{-4}  
\left( \frac{\epsilon_H}{\epsilon_R} \right)^3  \frac{e^{2\pi \xiend}}{\xiend^3} \, .
 \label{eq:DMeq}
\eea
The observed energy density of cold dark matter today~\cite{Ade:2015xua} is $\rho_{\rm CDM} = 9.6 \times 10^{-48} \ {\rm GeV}^4$.~The contribution to the cold dark matter relic density from the transverse dark photon mode produced via tachyonic instability can be written as $\Omega_T / \Omega_{\rm CDM} = \rho_D(T_0) / \rho_{\rm CDM}$ where,
\bea
\frac{\Omega_T}{\Omega_{\rm CDM}} &=& 
7 \times 10^{-6} \, \frac{m}{\rm GeV} 
\left( \frac{H}{10^{11} \ {\rm GeV}} \right)^{3/2}  \nn\\
&\times&
\left( \frac{\epsilon_H}{\epsilon_R} \right)^3  \frac{e^{2\pi \xiend}}{\xiend^3} \, .
\label{eq:DMT}
\eea
We see that this depends on five parameters:~the dark photon mass $m$, the Hubble scale during inflation $H$, $\epsilon_H$ which parametrizes how much $H$ has decreased by the end of inflation (see~\eref{Hend}), $\epsilon_R$ which parametrizes the fraction of energy density transferred from the inflaton to radiation at reheating (see~\eref{rhoRH}), and $\xiend$ which parametrizes the strength of the inflaton - dark photon coupling (see~\eref{xidef}) and implicitly depends on the form of the inflaton potential.

There is also a contribution to the relic density from the longitudinal mode which is produced via quantum fluctuations during inflation as described in~\cite{Graham:2015rva},
\bea \label{eq:DML}
\frac{\Omega_L}{\Omega_{\rm CDM}} = \left( \frac{m}{ 6 \times 10^{-15} \ {\rm GeV}} \right)^{1/2}  \left( \frac{H}{10^{14} \ {\rm GeV}} \right)^2 \, .~~~
\eea
To understand whether the transverse or longitudinal mode dominates the relic density, we will explore the parameter space in the $m - H$ plane, but first we must ensure the following constraints are satisfied for the tachyonically produced transverse dark photon mode:
\begin{enumerate}
\item The dark photon mass must be much smaller than $q(\TRH)$ to allow for efficient tachyonic production at the end of inflation.~At the same time $m$ is bound from below by the condition that the dark photon becomes non-relativistic, thus behaving like cold dark matter, before matter-radiation equality.~We write this condition as $\bar T > \TCMB$, with $\TCMB \simeq 10^{-9}$ GeV.~The dark photon mass is therefore constrained to be in the window,
\bea \label{eq:mconstraint}
\left(\frac{\pi g_*(\TRH)}{90}\right)^{1/4} \frac{\epsilon_H}{\epsilon_R} \sqrt{\frac{H}{\Mpl}} \, \TCMB < m \ll  \epsilon_H H \, .~
\eea
\item At reheating, the energy density of radiation in the visible sector must be greater than that in the dark photon sector, $\rho_R (\TRH) > \rho_D(\TRH)$ which implies,
\bea \label{eq:epsilonconstraint}
\left( \frac{\epsilon_H}{\epsilon_R} \right)^4 < 3 \times 10^4 \, \xiend^3 e^{-2\pi \xiend} \left( \frac{\Mpl}{H}  \right)^2 \, .
\eea
If this condition were not satisfied, the Universe would become (dark) matter dominated at a temperature $\bar T > \TCMB$, thus violating matter-radiation equality at $\TCMB$ as well as not allowing for enough time between $\bar T$ and $\TCMB$ for the dark photons to become cold (non relativistic). 
\item If the inflaton - dark photon coupling is too large, the dark photon can thermalize with the inflaton.~Since the latter must also couple to Standard Model particles to reheat the Universe, this would lead to thermalization of the dark photon with the visible sector and spoil our dark matter production mechanism.~Ensuring the inflaton and dark photon do not thermalize~\cite{Ferreira:2017lnd, Ferreira:2017wlx} puts an upper bound on $\xi$,
\bea
\xi < 0.44 \ln \frac{f}{\alpha H} + 3.4 \, .
\eea
Using~\eref{xidef} and taking the slow-roll parameter to be $\epsilon = 1$ at the end of inflation then gives,
\bea \label{eq:thermal}
\xiend < 0.44 \ln \left( \frac{1}{\sqrt{2} \xiend} \frac{\Mpl}{H}  \right) + 3.4 \, .
\eea
We must also ensure that there are no other light scalar or fermion fields in the dark sector which couple to the dark photon.~If such light fields were present, they would be produced by the strong dark electromagnetic field via the Schwinger effect~\cite{Tangarife:2017vnd, Tangarife:2017rgl} and the dark sector would thermalize, thus spoiling our production mechanism.
\item We also assume that back-reaction effects on the inflaton dynamics are negligible which leads to two conditions.~The first is $3 H \dot\phi \simeq V' \gg \alpha / f \langle \vec E \cdot \vec B \rangle$, meaning that the $\langle F\tilde{F}\rangle = \langle \vec E \cdot \vec B \rangle$ term is negligible in the inflaton equation of motion in~\eref{EOMphi}.~The second condition is that $3 H^2 \Mpl^2 \gg 1/2 \langle \vec E^2 \rangle$, meaning that the inflaton dominates the energy density during inflation rather than the dark photon.~Both conditions are satisfied as long as $\xi$ is not too large~\cite{Barnaby:2011qe,Peloso:2016gqs}.~Requiring that they hold all the way to the end of inflation results in the constraint,
\bea \label{eq:backreaction}
\frac{\epsilon_H H}{\Mpl} \ll 10^2\, \xiend^{3/2} \, e^{-\pi\xiend} \, .
\eea
\item When the inflaton exits the slow-roll regime, it starts oscillating about the minimum of its potential and reheats the Universe.~If the coupling $\alpha / f$ is moderately large, roughly $\alpha / f > 35 \Mpl^{-1}$, the production of dark photons during these oscillations can be important.~This phenomenon is referred to as gauge-preheating and has been studied in~\cite{Adshead:2015pva}.~However, in order to satisfy the constraints listed above, here we consider the range $1 \lesssim \xi_{\rm{end}} \lesssim 10$.~This results in the range,
\bea
\frac{\sqrt{2}}{M_{\rm{Pl}}} \lesssim \frac{\alpha}{f} \lesssim 
\frac{10 \sqrt{2}}{M_{\rm{Pl}}}  \, ,
\eea
where we have used the relation between $\alpha / f$ and $\xiend$ obtained from~\eref{xidef} and setting $\epsilon =1$ at the end of inflation
\footnote{Note, that in hybrid inflation models~\cite{Copeland:1994vg,Dvali:1994ms} one can have a slow-roll parameter different from $\epsilon \approx 1$ at the end of inflation.}.
In this range of $\alpha / f$, preheating into dark photons is not efficient~\cite{Adshead:2015pva} implying that during the inflaton oscillations only a negligible fraction of its energy density is transferred to the dark photon.~We then assume that reheating proceeds via the perturbative decay of the inflaton into Standard Model particles and approximate the process as instantaneous. 
\end{enumerate}
In~\fref{relic} we show the main result of our study given in~\eref{DMT}, imposing the constraints listed above, for different values of the parameters $\xiend$, $\epsilon_R$, $\epsilon_H$.~For comparison, we also plot the relic abundance of the longitudinal mode given in~\eref{DML} and produced via inflationary fluctuations~\cite{Graham:2015rva}.~In the regions where the line labelled `Transverse' is to the left of the one labelled `Longitudinal', the transverse mode gives the dominant contribution to the dark matter relic density.~We see there are large regions of parameter space where this is the case.
%%%%%%%%%%%%%%%%
%%%%%
\begin{figure*}
\includegraphics[width=6cm]{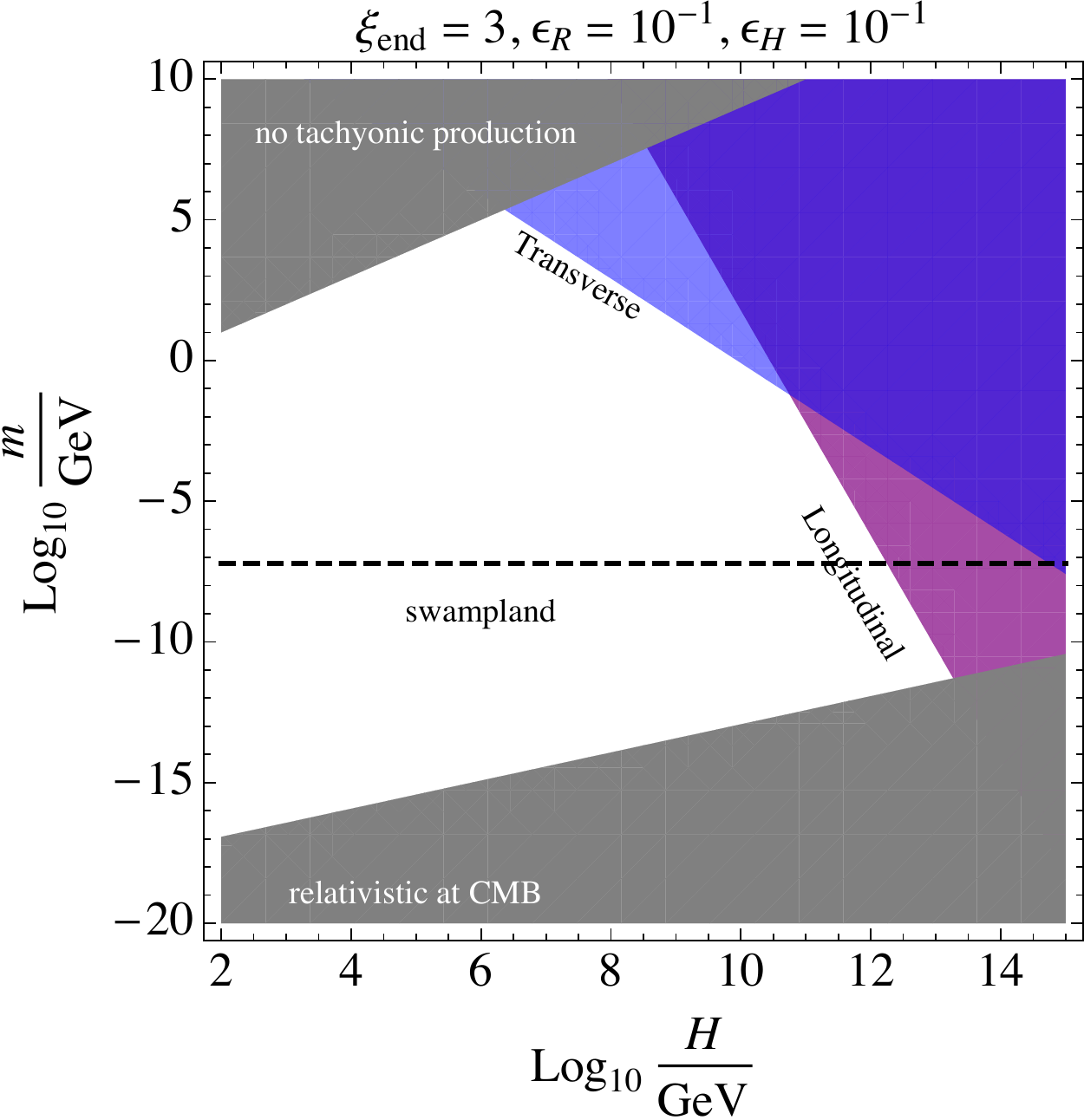}
\hspace{1cm}
\includegraphics[width=6cm]{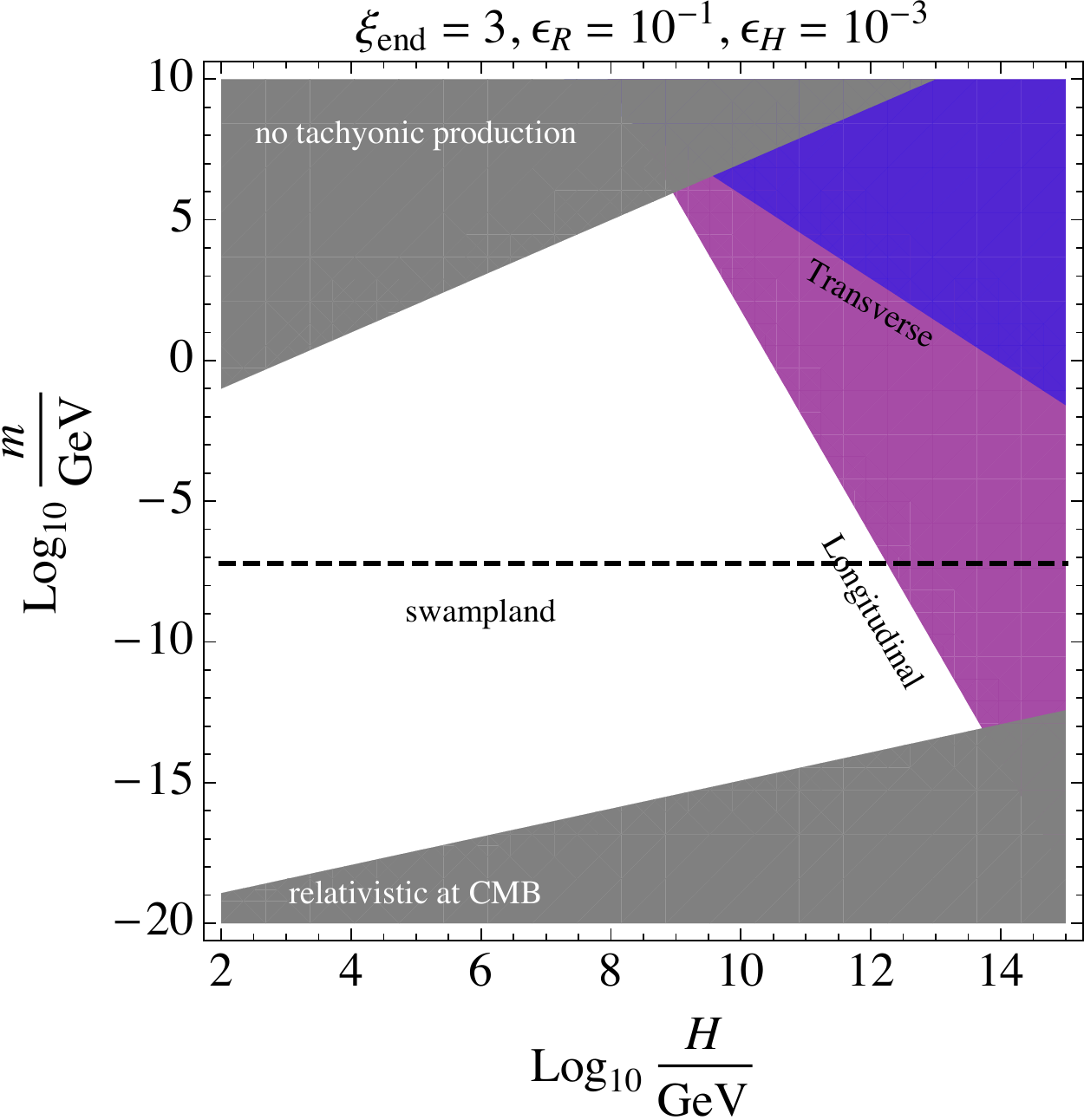}\\
\vspace{0.5cm}
\includegraphics[width=6cm]{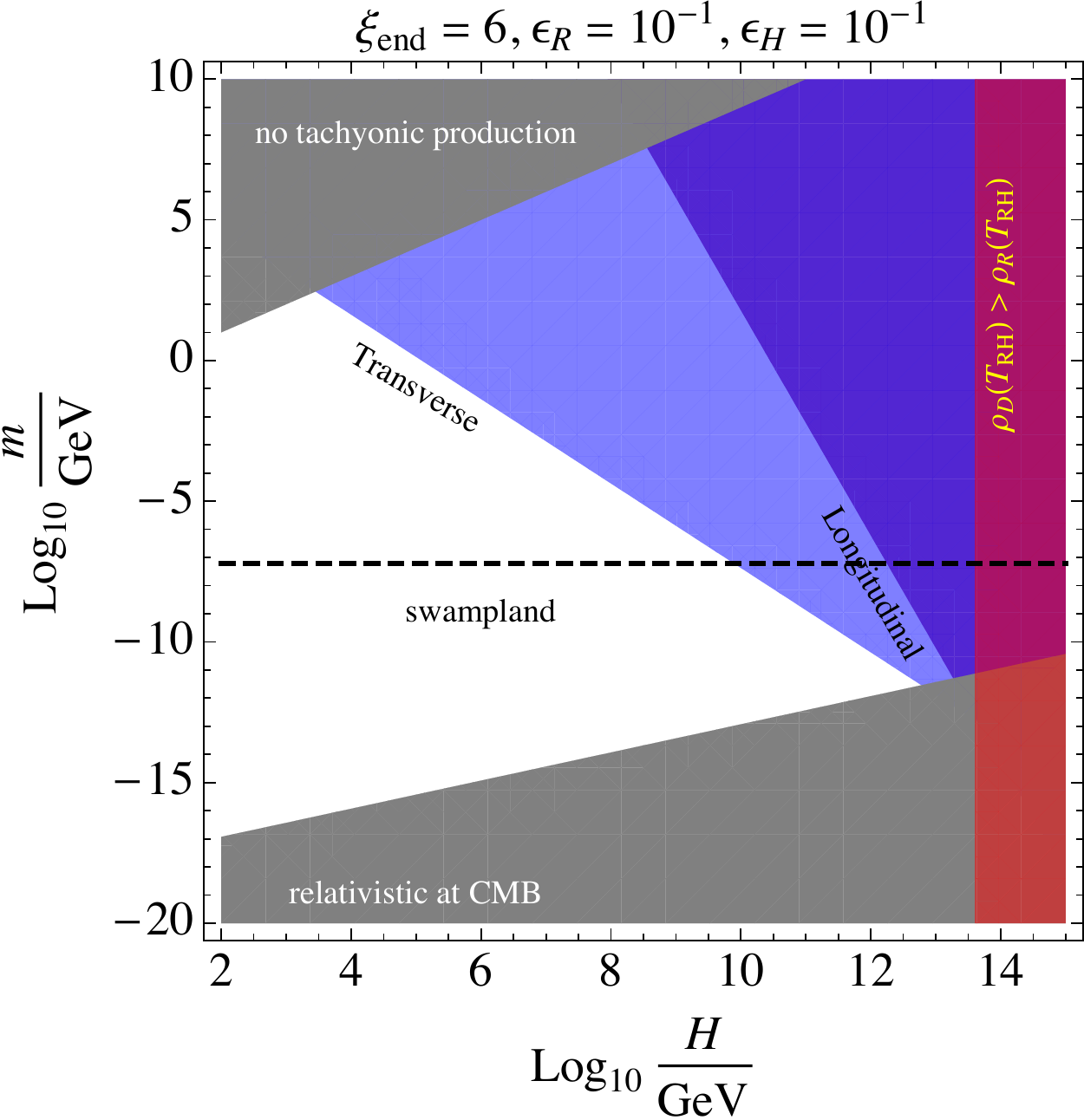}
\hspace{1cm}
\includegraphics[width=6cm]{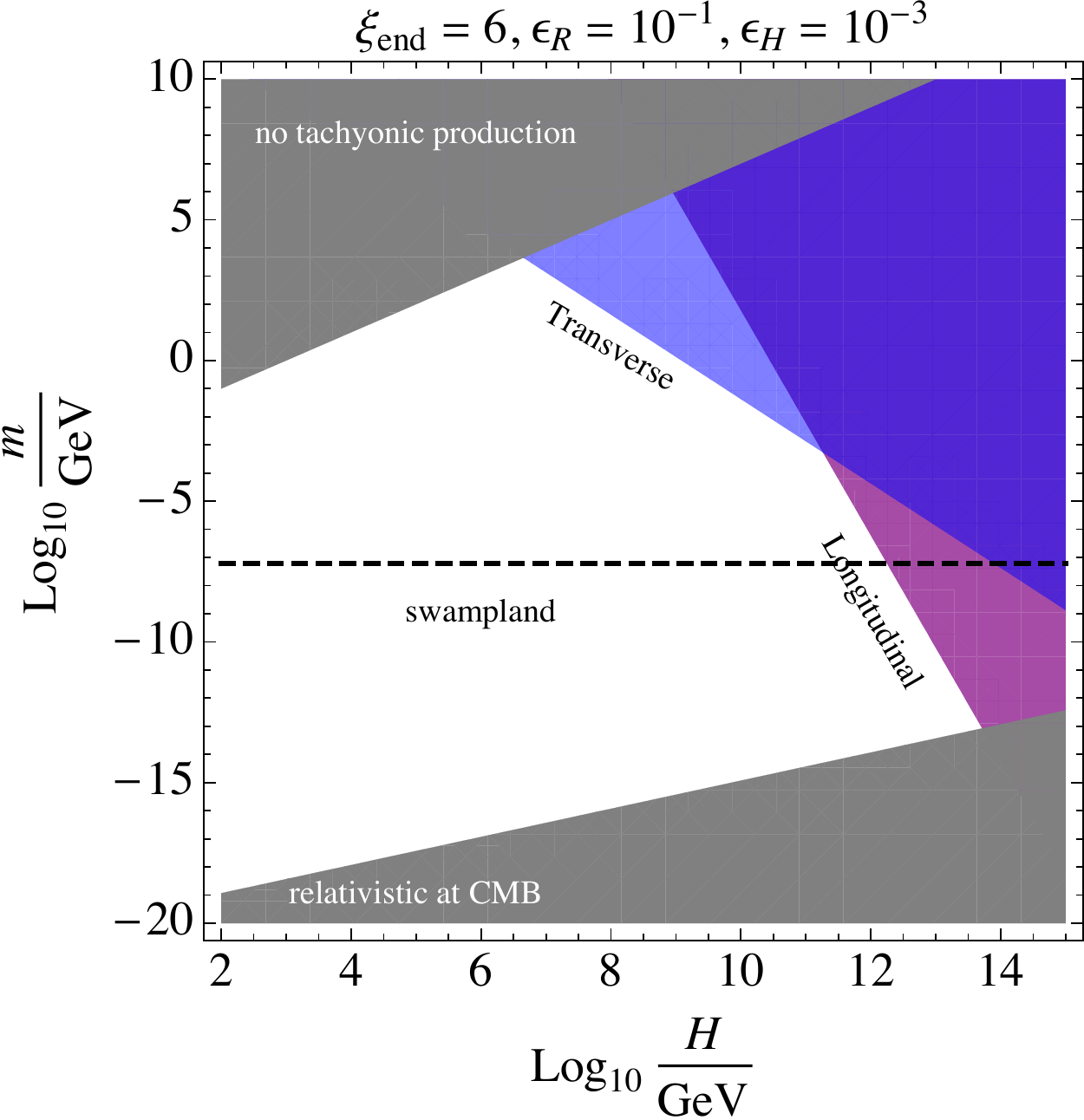} \\
\vspace{0.5cm}
\includegraphics[width=6cm]{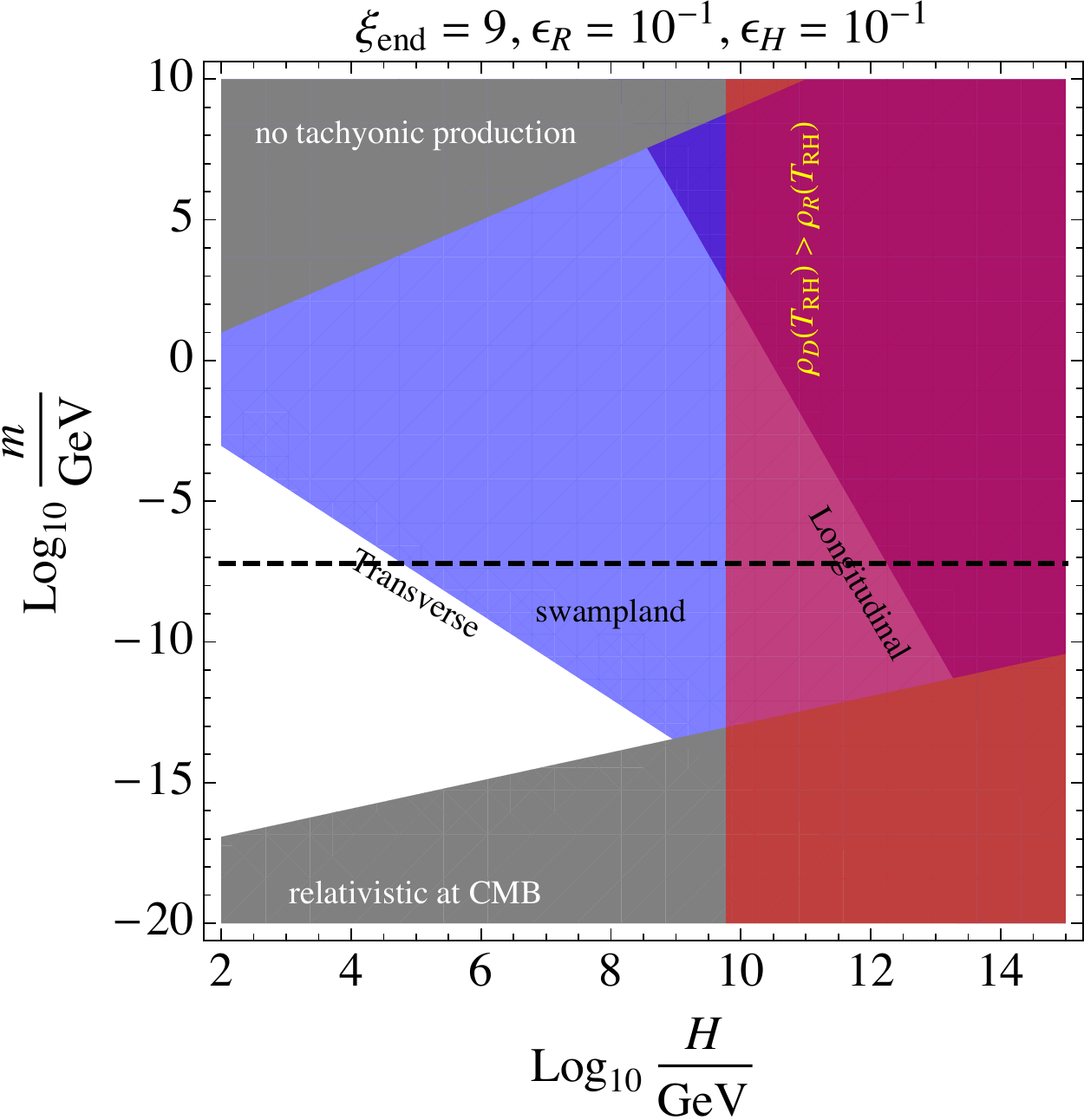}
\hspace{1cm}
\includegraphics[width=6cm]{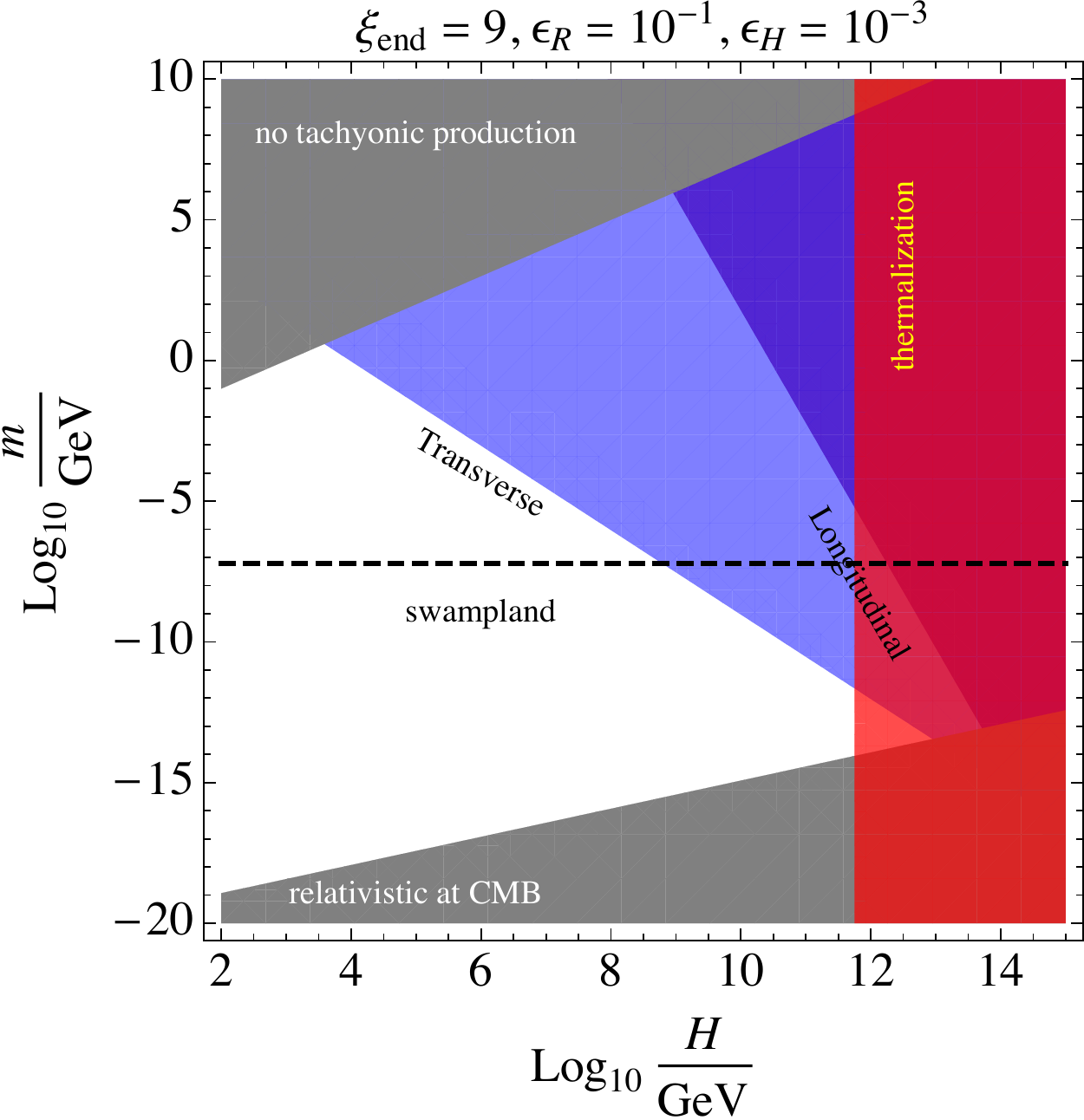}
\caption{Parameter space in the dark photon mass versus Hubble scale $m - H$ plane for values of the parameters $\xiend$, $\epsilon_R$, and $\epsilon_H$ as indicated above each plot and described in the text.~Along the blue (purple) line, next to the label ``Transverse'' (``Longitudinal''), we obtain the observed relic abundance for the transverse (longitudinal) mode while to the right of these lines the dark matter is overabundant.~The region in the gray band at large masses is excluded by requiring efficient tachyonic production during inflation while the region in gray at low masses is excluded by requiring the dark photons are non-relativistic by the time of CMB formation (see~\eref{mconstraint}).~The regions in the red vertical bands, present only for $\xiend=6,9$, are excluded by the strongest among the constraints in~\eref{epsilonconstraint}, \eref{thermal}, and \eref{backreaction}.~Specifically, the region labeled `$\rho_D(\TRH) > \rho_R(\TRH)$' is excluded by the constraint in~\eref{epsilonconstraint} and the region labeled `thermalization' is excluded by~\eref{thermal}.~Below the dashed horizontal line, the production of the longitudinal mode is in conflict with constraints from conjectures of quantum gravity~\cite{Reece:2018zvv}.}
\label{fig:relic}
\end{figure*}
%%%%%
%

Finally, we consider a specific benchmark point,
\bea\label{eq:bench}
\xiend=6\, , \ \ \epsilon_R = 10^{-1}\, , \ \ \epsilon_H=10^{-3}\, , \nn \\ 
H = 10^9 \ \Gev\, , \ \ m = 1.3 \ \Gev\, .
\label{eq:bench}
\eea
This leads to a reheating temperature $\TRH = 2.7 \times 10^{12}$ GeV and an initial radiation energy density $\rho_R(\TRH) = 1.7 \times 10^{51} \ \Gev^4$, several orders of magnitude larger than the initial energy density in the dark electromagnetic field $\rho_D(\TRH) = 10^{34} \ \Gev^4$.~The dark photons become non-relativistic at $\bar T = 3.6 \times 10^6$ GeV, then redshift like matter for some time before matching the energy density of radiation at $\TCMB$, thus giving a viable dark matter candidate for this benchmark point.~Note that the momentum of the dark photon has a long time to redshift from $\bar T$ to $\TCMB$ so it is very `cold' by the time of matter-radiation equality.~Note also that at the time of Big Bang Nucleosynthesis (BBN), $T_{\rm BBN} \sim 1$ MeV, the dark photon is already non-relativistic and still constitutes a small fraction of the total energy density.~Therefore there are no bounds on this scenario from considerations on extra relativistic species ($N_{\rm eff}$).  

%%%%%%%%%%%%%%%%%%%%%%%%%%%%%%%%%%%%%%%%%%%%%
\section{Stueckelberg mass in the swampland} \label{sec:swampland}

In our model defined by the Lagrangian in~\eref{Lag}, we have implicitly included a Stueckelberg mass for the dark photon.~However, as discussed this assumption is not necessary for the mechanism described here since it produces only the transverse mode.~If instead the dark photon obtains its mass from a Higgs mechanism via a coupling to a scalar which obtains a vacuum expectation value (\emph{vev}), the transverse modes are well defined whether in the broken (non-zero \emph{vev}) or unbroken phase (zero \emph{vev}).~The tachyonic production mechanism during inflation applies in either case assuming the mass is small compared to the Hubble scale in the Stueckelberg case.~The only requirement is that during the radiation dominated era the dark photon either already has a mass, as in the Stueckelberg case, or becomes massive after a phase transition from zero to non-zero \emph{vev} as in the Higgs-ed case.~In either case our analysis is unaffected since the vector becomes non-relativistic when its typical momentum redshifts to a value below its mass, regardless of its origin. 

This is in contrast to the longitudinal mode since, as explained in~\cite{Graham:2015rva}, the production via inflationary fluctuations relies on the fact that the Stueckelberg mass is the correct effective description all the way up to the potentially very high scale of inflation.~Whether the Stueckelberg description is valid up to an arbitrarily high cutoff and whether this is consistent with theories of quantum gravity has been the subject of recent investigation~\cite{Reece:2018zvv}.~While the arguments are not based on a rigorous proof, they provide some sort of theoretical guidance.~It was found that for a spin-1 vector boson with a Stueckelberg mass, the theory breaks down at a cutoff of $\Lambda_{\rm UV} =  e^{1/3} \Mpl$ where $e$ is the gauge coupling.

Applying this conjecture to the mechanism of~\cite{Graham:2015rva} and requiring the Hubble scale of inflation to be below $\Lambda_{\rm UV}$, one finds the constraint $m > 60$ eV (see Eq.~(29) in~\cite{Reece:2018zvv} and related discussion for more detail).~For masses below this value, which we plot for reference in~\fref{relic} (dashed line), the production mechanism for the longitudinal mode seems to be inconsistent with quantum gravity and lives in the swampland (see~\cite{Craig:2018yld} for a way this bound might be evaded).~Of course since the mechanism proposed in this work does not rely on a Stueckelberg mass it is not (necessarily) subject to this constraint.

%%%%%%%%%%%%%%%%%%%%%%%%%%%%%%%%%%
\section{Summary and outlook} \label{sec:const}

We have presented a new mechanism for generating vector dark matter at the end of inflation.~It relies on the coupling of the inflaton to the vector which leads to exponential production of one of the transverse polarizations of the dark photon.~The main results are given~\eref{DMT} and~\fref{relic} where we show the parameter space for the dark photon relic abundance.~We find that the dark photon can make a viable dark matter candidate over a wide range of parameter space:
$\mu {\rm eV} \lesssim m \lesssim {\rm TeV}$,
$100 \ {\rm GeV} \lesssim H \lesssim 10^{14} \ {\rm GeV}$,
$\sqrt{2} \Mpl^{-1} \lesssim \alpha / f \lesssim 10\sqrt{2} \Mpl^{-1}$.

We have not specified the inflaton potential since its precise form is not crucial for the tachyonic production mechanism.~We have also assumed a small enough inflaton - dark photon coupling that back-reaction effects on the inflaton dynamics are negligible as well as instantaneous reheating at the end of inflation.~Even under these constraining assumptions, we have found a large region of parameter space in which this mechanism can produce the observed dark matter relic density. 

There are several interesting future directions to pursue.~A detailed study of the energy density spectrum, tracking its evolution from inflation to the present time, is essential for a more a more precise quantification of the relic density.~It can also provide important information on the length scales of dark matter structures~\cite{Graham:2015rva,Alonso-Alvarez:2018tus} which can be relevant for dark matter search strategies.~A detailed study of the power spectrum is in progress~\cite{followup} and will be presented in the near future.~One can also relax some of our assumptions and see if more parameter space opens up.~This requires, in particular, a detailed study of the preheating phase where the inflaton undergoes oscillations.~This can also result in significant dark photon production for large values of its coupling to the inflaton as has been shown recently~\cite{Agrawal:2018vin,Co:2018lka} for an oscillating axion.~Vector production can also occur during the oscillations of a dark Higgs which gives a mass to the dark photon~\cite{Dror:2018pdh}.~This type of dark Higgs mechanism can also be incorporated into the scenario proposed here and lead to potentially interesting gravitational wave signals~\cite{Breitbach:2018ddu}.~Furthermore, allowing for non-zero (but small) kinetic mixing between the dark and visible photons can lead to interesting dark matter phenomenology~\cite{Chaudhuri:2014dla}.~Another potentially interesting future direction is the fact that only one transverse polarization is produced in this mechanism.~This could lead to unique features associated with the polarization which can be searched for experimentally~\cite{Alonso:2018dxy,Wolf:2018xlz}, but is left for future investigation.

%%%%%%%%%%%%%%%%%%%%%%%%%%%%%%%%%%%%%%

\noindent
{\bf Acknowledgments:}~We thank Diego Blas,~Bohdan Grzadkowski,~Takeshi Kobayashi,~Gilad Perez,~Pedro Schwaller,~Javi Serra,~Tomer Volansky and Tien-Tien Yu for useful comments and discussions.~This work has been partially supported by MINECO grants~FPA 2016-78220-C3-1-P,~including ERDF (J.S.,\,R.V.M.),~FIS2016-7819-P (M.B.G.),~and Junta de Andaluc\'{i}a Project FQM-101 (M.B.G.,\,J.S.,\,R.V.M.), as well as the Juan de la Cierva program (R.V.M.).~L.U.~acknowledges support from the PRIN project ``Search for the Fundamental Laws and Constituents'' (2015P5SBHT\textunderscore 002).
%\clearpage 
%

\bibliographystyle{apsrev}
\bibliography{references}

\end{document}